\documentclass[twocolumn,showpacs,superscriptaddress,
preprintnumbers,floatfix]{revtex4}
\usepackage{graphicx,epsf,amssymb,amsmath,latexsym}
\newcommand{\be}{\begin{eqnarray}}
\newcommand{\ee}{\end{eqnarray}}

\renewcommand{\d}{\mbox{${\rm d}$}}

\begin{document}
\title{The Method of Comparison Equations for Schwarzschild Black Holes}
\author{Roberto Casadio}
\email{Roberto.Casadio@bo.infn.it}
\affiliation{Dipartimento di Fisica, Universit\`a di Bologna and I.N.F.N.,
Sezione di Bologna, via~Irnerio~46, 40126 Bologna, Italy.}
\author{Mattia Luzzi}
\email{Mattia.Luzzi@bo.infn.it}
\affiliation{Dipartimento di Fisica, Universit\`a di Bologna and I.N.F.N.,
Sezione di Bologna, via~Irnerio~46, 40126 Bologna, Italy.}
\begin{abstract}
We employ the method of comparison equations to study the propagation
of a massless minimally coupled scalar field on the Schwarzschild background.
In particular, we show that this method allows us to obtain explicit approximate
expressions for the radial modes with energy below the peak of the effective potential
which are fairly accurate over the whole region outside the horizon.
This case can be of particular interesting, for example,
for the problem of black hole evaporation.
\end{abstract}
\maketitle
\section{Introduction}
Wave equations on black hole backgrounds cannot be solved exactly
in terms of known simple functions (for a review, see, {\em e.g.}, Ref.~\cite{chandra}).
Although it is usually fairly easy to find asymptotic expressions near the (outer)
horizon or at spatial infinity, numerical analysis is what one eventually employs
to obtain a detailed profile of the wave-functions.
However, our understanding of the field dynamics on curved backgrounds
would still benefit of analytical approximations to the exact modes,
for example in the study of the (renormalized) energy-momentum tensor of quantum
fields~\cite{birrell} and, in particular, of the Hawking radiation~\cite{hawking},
or of quasi-normal modes~\cite{nollert}.
\par
By making use of the black hole symmetry, one can usually separate the wave-functions
and reduce the problem to the one task of solving a second order differential equation
for the radial part.
The latter takes the form of the one-dimensional Schr\"odinger equation for the
transmission through a potential barrier~\cite{chandra}
\be
\left[\frac{\d^2}{\d x^2}+Q_s(x)\right]\psi=0
\ ,
\label{gen}
\ee
and, for the general case of a Kerr-Newman black hole, is known as the Teukolsky
master equation for waves of spin-weight $s=0$, $1$ and $2$~\cite{teukolsky,chandra}.
\par
A well known method to solve equations of the form~(\ref{gen}) is the Wentzel-Kramers-Brillouin
(WKB) approximation~\cite{wkb,langer}, which has in fact been widely used to study
wave equations on black hole backgrounds (for one of the first attempts see Ref.~\cite{iyer}
and for recent reviews see, {\em e.g.}, Ref.~\cite{grain} and References therein).
One of the weaknesses of this method is that it becomes rather cumbersome near the
``turning points'', that is where the potential vanishes and one must match
different branches of approximate solutions.
The precise location of the matching points greatly affects the accuracy of
the method and may lead to errors at next-to-leading order larger that those
at leading order (for a detailed analysis of this point in the context of
cosmology, see Ref.~\cite{impWKBprd}).
The fact that the potentials in the radial equations on black hole backgrounds
can in general have two turning points makes therefore it difficult to apply the
WKB approximation.
\par
There are however improved versions of the WKB approximation and, in the
present paper, we shall concentrate on the one which seems to give
the most accurate leading order expressions.
The method of comparison equations (MCE) was independently introduced in
Refs.~\cite{MG,dingle} and later applied, for example, to wave
mechanics~\cite{berry} and cosmology~\cite{MCEcosmo}.
Its connection with the Ermakov-Pinney equation was also studied in
Ref.~\cite{KLV}.
Its strength consists in yielding approximate modes which are rather
accurate over the entire domain of definition of the wave-functions.
The approximate wave-functions are however given in terms of a new
independent variable and computing their explicit dependence on
the original variable is now the most difficult task which requires
a good deal of trial-and-error work.
\par
In the next Section, we shall briefly review the equation of motion
for massless scalar fields on a spherically symmetric black hole
background.
This includes the simplest case of linear scalar perturbations of a
Schwarzschild black hole which will be used to show the
application of our method in all details (including the use of a convenient
radial coordinate and rescaling of the wave-function).
In Section~\ref{sMCE_BH}, we shall then apply the MCE and obtain
approximate expressions for scalar wave-modes with energy below the
peak of the potential.
This case is of particular interest, for example, for evaluating the grey-body
factors involved in the Hawking effect~\cite{hawking}.
It also presents a major complication with respect to the typical cases treated
in Ref.~\cite{MCEcosmo}, namely the presence of two zeros of the
potential (``turning points'').
We shall use units with $c=G=1$.
\section{Scalar field on Schwarzschild background}
\label{sSBH}
Let us begin by recalling that the metric in a static and spherically
symmetric vacuum space-time can be written in general as
\be
\d s^2=h(r)\,\d t^2-\frac{\d r^2}{h(r)}-r^2\,\d\Omega^2
\ ,
\label{metric}
\ee
where $\d\Omega$ is the area element on the unit two-sphere.
The propagation of massless, minimally coupled scalar particles in this
background is governed by the Klein-Gordon equation
$\Phi^{;\mu}_{\ \ ;\mu}=0$.
Due to the symmetry of the background, the field $\Phi$ can be decomposed into
eigenmodes of normal frequency $\bar\omega$ and angular momentum
numbers $\ell$, $m$ as~\cite{chandra}
\be
\Phi(t,r,\Omega)={\rm e}^{-i\,\bar\omega\,t}\,Y^{m}_{\ell}(\Omega)\,R(r)
\ ,
\ee
where $Y_\ell^m$ are spherical harmonics.
The Klein-Gordon equation then separates and the dynamics
is described by the function $R$ which satisfies the radial equation
\be
\frac{\d}{\d r}\left[r^2\,h(r)\,\frac{\d R(r)}{\d r}\right]
+\left[\frac{\bar\omega^2\,r^2}{h(r)}-\ell\,(\ell+1)\right]R(r)=0
\ ,
\label{rad_eq}
\ee
where $\ell\,(\ell+1)$ is the separation constant.
No exact solution of the above equation is known even for as simple a background
as the four-dimensional Schwarzschild space-time with
\be
h(r)=1-\tilde{r}^{-1}
\ ,
\label{h_schw}
\ee
where we have introduced the dimensionless radial coordinate
$\tilde r=r/r_{\rm H}$ and $r_{\rm H}$ is the Schwarzschild radius.
\par
\begin{figure}[h]
\includegraphics[width=0.45\textwidth]{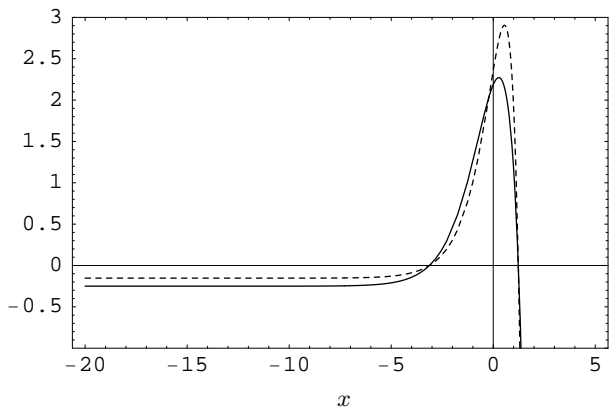}
\\
{\hspace{1cm}$x$}
\caption{Comparison between the exact potential $-\omega^2(x)$ (solid line)
and the approximate Morse potential $-\Theta^2[\sigma(x)]$ obtained after
solving Eqs.~(\ref{A_B}) and~(\ref{third_cond}) (dotted line) for
$\tilde\omega=1/2$ and $\ell=2$.
The two turning points are $x_1\approx -3.13$ and $x_2\approx 1.23$.}
\label{plot1}
\end{figure}

It is now convenient to introduce the ``tortoise-like''
coordinate~\footnote{A similar coordinate was already used
in Ref.~\cite{kanti_russ}.}
\begin{subequations}
\be
\d\,x=\frac{\d\,\tilde r}{\tilde r\,h(\tilde r)}
\ ,
\label{new_var}
\ee
and the new radial function
\be
R\left[\tilde r(x)\right]
={\rm exp}\left[-\frac12\,\int^{x}\,h(x')\,\d\,x'\right]\chi\left[\tilde r(x)\right]
\ ,
\label{new_funct}
\ee
\end{subequations}
so that the radial equation takes the Schr\"odinger form
\begin{widetext}
\be
\frac{\d^2\,\chi(x)}{\d x^2}
+\left\{\tilde\omega^2\,\tilde r^2(x)-\ell\,(\ell+1)\,h[\tilde r(x)]
-\frac{1}{4}\,h^2[\tilde r(x)]-\frac{1}{2}\,\frac{\d h[\tilde r(x)]}{\d x}\right\}\chi(x)=0
\ .
\label{new_rad_eq}
\ee
\end{widetext}
where we also defined the dimensionless energy
$\tilde{\omega}\equiv\bar\omega\,r_{\rm H}$.
For the four-dimensional Schwarzschild case we have
\be
x(\tilde{r})=\ln (\tilde{r}-1)
\ ,
\label{hatr_r}
\ee
so that the new variable $x$ ranges from
$x(\tilde r_{\rm H}=1)=-\infty$ to $x(\tilde r=+\infty)=+\infty$
and $x$ and $\chi(x)$ are of Langer's form~\cite{langer}.
The Klein-Gordon equation finally becomes
\be
\left[
\frac{{\d}^2}{{\d}x^2}+\omega^2(x)
\right]
\,\chi(x)=0
\ ,
\label{exact_EQ}
\ee
where the ``frequency''
\be
\omega^2(x)
=
\tilde{\omega}^2\left(1+{\rm e}^{x}\right)^2
-\frac{\ell\,(\ell+1)\,{\rm e}^{x}}{1+{\rm e}^{x}}
-\frac{{\rm e}^{x}\left(2+{\rm e}^{x}\right)}{4\,\left(1+{\rm e}^{x}\right)^2}
\ ,
\label{freq_ex}
\ee
is not necessarily a positive quantity.
In fact, the above expression shows two qualitatively
different behaviors depending on the values of $\tilde{\omega}$
and $\ell$.
In particular, for a given angular momentum $\ell$, there exists
a critical value
\be
\tilde\omega_{\rm c}=\tilde\omega_{\rm c}(\ell)
\ ,
\label{om_c}
\ee
such that if the energy $\tilde{\omega}>\tilde\omega_{\rm c}$ there are
no turning points, otherwise there are two, say
$\omega(x_1)=\omega(x_2)=0$ with $x_1<x_2$.
The latter case is the one we want to analyze in detail
(see Appendix~\ref{app_TPs} for more details and
the solid line in Fig.~\ref{plot1} for an example).
\section{MCE solutions}
\label{sMCE_BH}
The MCE makes use of the exact solutions to a
similar second-order differential equation,
\be
\left[
\frac{{\d}^2}{{\d}\bar\sigma^2}+\Theta^2(\bar\sigma)
\right]
U(\bar\sigma)=0
\ ,
\label{aux_EQ}
\ee
where $\Theta$ is the ``comparison function''.
One can then represent exact solutions of Eq.~(\ref{exact_EQ})
in the form
\be
\chi(x)=\left(\frac{{\d}\bar\sigma}{{\d}x}\right)^{-1/2}
U(\bar\sigma)
\ ,
\label{exact_SOL}
\ee
provided the variables $\bar\sigma$ and $x$ are related by
\be
\omega^2(x)\!=\!\left(\frac{{\d}\bar\sigma}{{\d}x}\right)^{2}\Theta^2(\bar\sigma)
-\left(\frac{{\d}\bar\sigma}{{\d}x}\right)^{1/2}
\frac{{\d}^2}{{\d}x^2}\left(\frac{{\d}\bar\sigma}{{\d}x}\right)^{-1/2}
.
\label{new_EQ}
\ee
Of course, solving Eq.~(\ref{new_EQ}) is usually as difficult as the original
problem~(\ref{exact_EQ}), and the dependence of $\bar\sigma$ on $x$ can just be
determined by using iterative schemes, in general cases~\cite{KLV,hecht}
or for specific problems~\cite{mori,pechukas}.
However, if we are able to find a comparison function $\Theta$ sufficiently similar to
$\omega$ in Eq.~(\ref{freq_ex}), the second term in the right hand side of
Eq.~(\ref{new_EQ}) will be negligible with respect to the first
one, so that
\be
\omega^2(x)\simeq\left(\frac{{\d}\bar\sigma}{{\d}x}\right)^{2}
\Theta^2(\bar\sigma)
\ .
\label{new_EQ_appr}
\ee
On selecting a pair of values $x_0$ and $\bar\sigma_0$ such that
$\bar\sigma_0=\bar\sigma(x_0)$, the function $\bar\sigma(x)$ can then be
approximated by a solution $\sigma=\sigma(x)$ of the
integral equation
\be
\int_{x_0}^x\sqrt{\pm\,\omega^2(y)}\,{\d}\,y
=
\int_{\sigma_0}^{\sigma}\sqrt{\pm\,\Theta^2(\rho)}\,{\d}\,\rho
\ ,
\label{new_EQ_int}
\ee
where the signs must be chosen conveniently.
This procedure leads to the MCE approximation for $\chi(x)$,
\be
\chi_{\rm MCE}(x)=\left(\frac{{\d}\sigma}{{\d}x}\right)^{-1/2}
U(\sigma)
\ ,
\label{chiMCE}
\ee
which is valid in the whole range of $x$ and including
the turning points.
\par
We are now dealing with a problem of the form~(\ref{exact_EQ})
with two turning points, $x_1<x_2$.
In order to implement the MCE, we need a ``comparison frequency''
with the same behavior.
We shall use the Morse potential~\cite{RMNS}
\be
\Theta^2(\sigma)=
A\,{\rm e}^{2\,\sigma}
-B\,{\rm e}^{\sigma}
+D
\ .
\label{comp_freq}
\ee
The coefficients $A$ and $B$ are then fixed by imposing
that the turning points of the comparison function are
the same as those of the exact frequency, that is
\begin{subequations}
\be
\Theta(\sigma_1=x_1)=\Theta(\sigma_2=x_2)=0
\ ,
\label{A_B}
\ee
and the coefficient $D$ will then follow form the relation
\be
\xi\equiv
\int_{x_1}^{x_2}\sqrt{-\,\omega^2(y)}\,{\d}\,y
=
\int_{x_1}^{x_2}\sqrt{-\,\Theta^2(\rho)}\,{\d}\,\rho
\ .
\label{third_cond}
\ee
\end{subequations}
The system of non-linear equations~(\ref{A_B}) and (\ref{third_cond}) thus
yields
\begin{subequations}
\be
A&\!=\!&
\frac{4\,\xi^2}{\pi^2\left({w_1+w_2}-2\,\sqrt{w_1\,w_2}\right)^2}
\label{coeff_A}
\\
B&\!=\!&
\left(w_1+w_2\right)A
\label{coeff_AB}
\\
D&\!=\!&
w_1\,w_2\,A
\ ,
\label{coeff_D}
\ee
\end{subequations}
where the exact expression for $\xi$ is given in Appendix~\ref{app_xi}.
A sample plot of both the exact frequency and the comparison function
thus obtained is given in Fig.~\ref{plot1}.
\par
The ``comparison solution'' is a linear combination of confluent hypergeometric
functions~\cite{abramowitz} of the type ${}_1{\rm F}_1$ (for more details,
see Ref.~\cite{RMNS}),
\begin{widetext}
\be
U(\sigma)&\!=\!&
C_+\,{\rm e}^{i\,\left(\sqrt{A}\,{\rm e}^{\sigma}+\sqrt{D}\,\sigma\right)}\,
{}_1{\rm F}_1
\left(\frac12+i\,\sqrt{D}+i\,\frac{B}{2\,\sqrt{A}},1+2\,i\,\sqrt{D},
-2\,i\,\sqrt{A}\,{\rm e}^\sigma\right)
\nonumber
\\
&&
+C_-\,{\rm e}^{i\,\left(\sqrt{A}\,{\rm e}^{\sigma}-\sqrt{D}\,\sigma\right)}\,
{}_1{\rm F}_1
\left(\frac12-i\,\sqrt{D}+i\,\frac{B}{2\,\sqrt{A}},1-2\,i\,\sqrt{D},
-2\,i\,\sqrt{A}\,{\rm e}^\sigma\right)
\label{sol_morse}
\ee
and $\sigma(x)$ is given implicitly by Eq.~(\ref{new_EQ_int}) with
\be
\int_{x_1}^{\sigma}\sqrt{\Theta^2(\rho)}\,{\d}\,\rho
=
\sqrt{\Theta^2(\sigma)}
-\ln\left[
\left(2\,A\,{\rm e}^\sigma-B
+2\,\sqrt{A\,\Theta^2(\sigma)}
\right)^{\frac{B}{2\,\sqrt{A}}}
\left(2\,D\,{\rm e}^{-\sigma}-B
+2\,{\rm e}^{-\sigma}\,\sqrt{\,D\,\Theta^2(\sigma)}
\right)^{\sqrt{D}}
\right]
\ .
\label{sigma_of_x}
\ee
\end{widetext}
\begin{figure*}[ht]
\raisebox{3.5cm}{$\sigma$}
\includegraphics[width=0.45\textwidth]{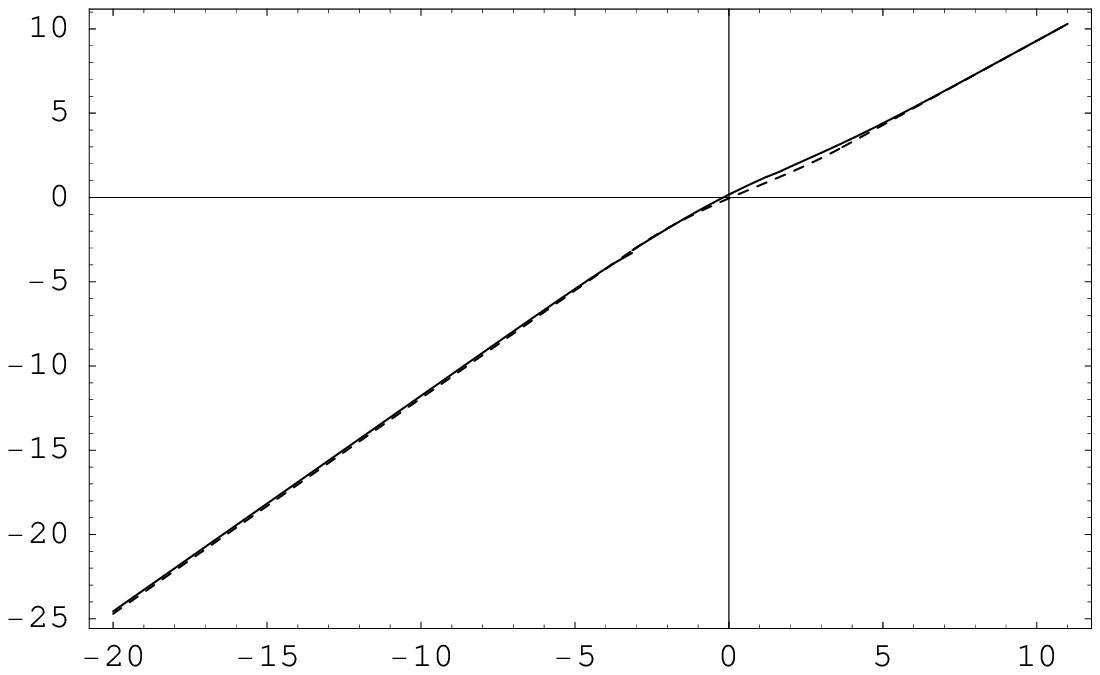}
\ \
\raisebox{3.5cm}{$\chi$}
\includegraphics[width=0.45\textwidth]{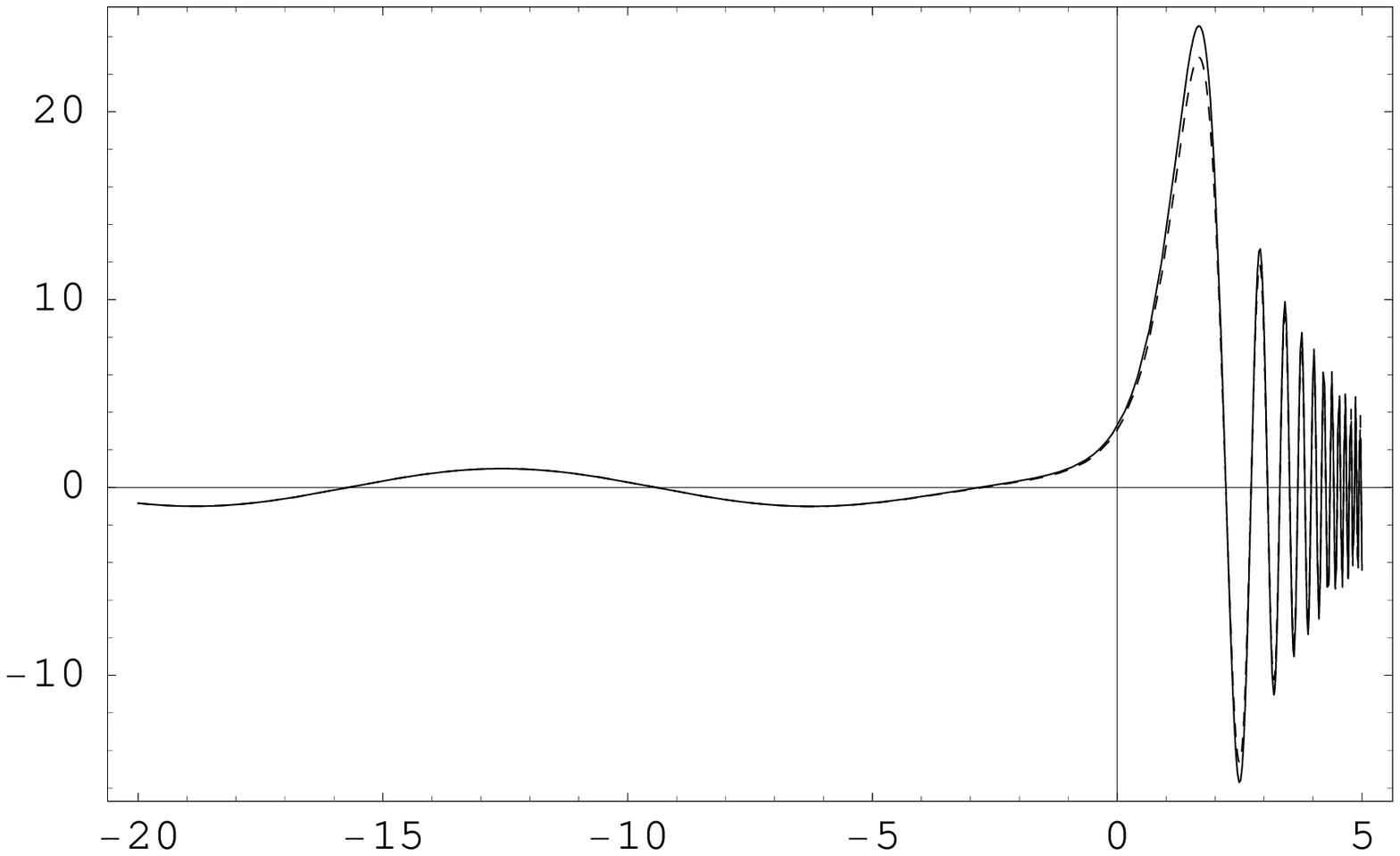}
\\
\hspace{1cm}$x$\hspace{8cm}$x$
\\
\vspace{0.5cm}
\includegraphics[width=0.32\textwidth]{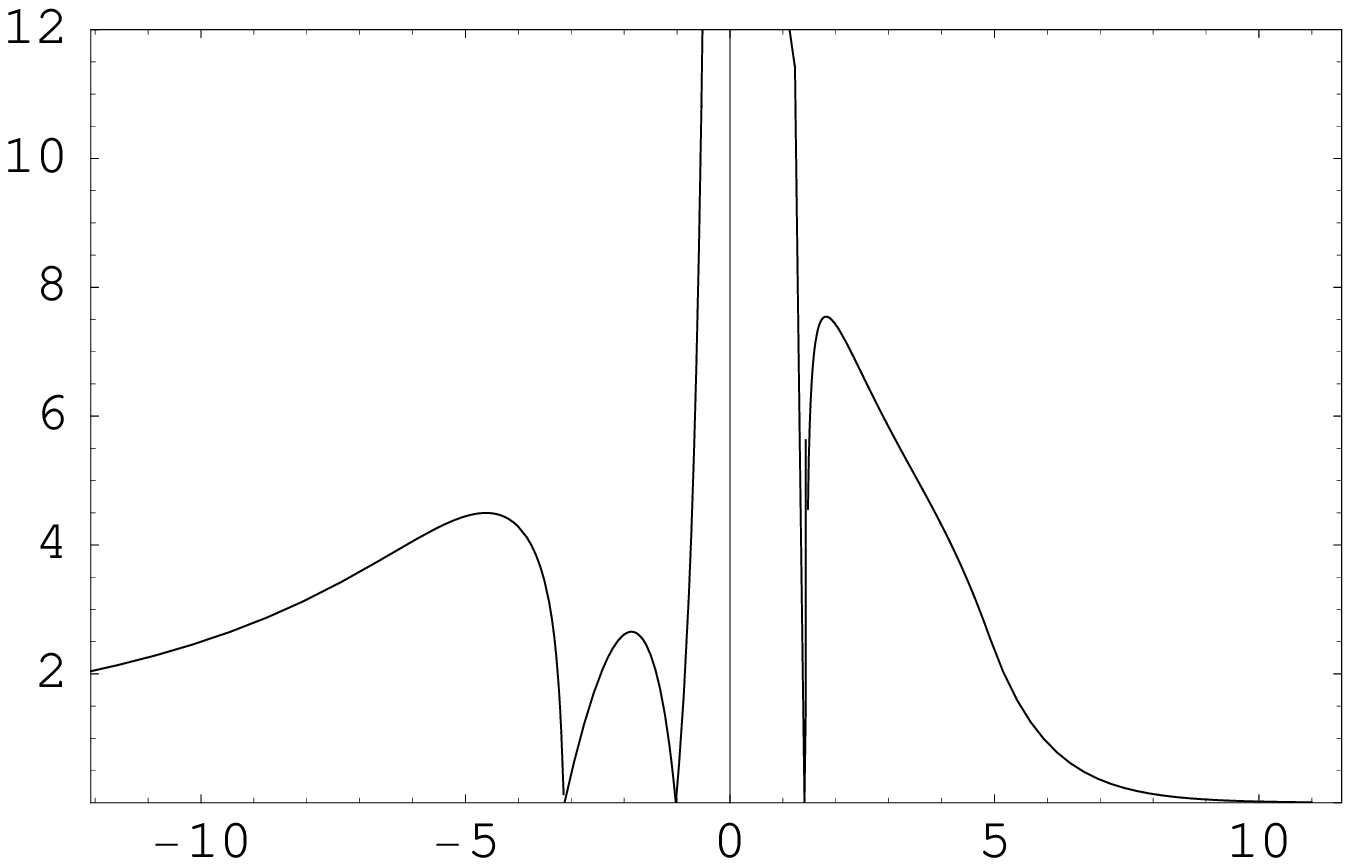}
\includegraphics[width=0.32\textwidth]{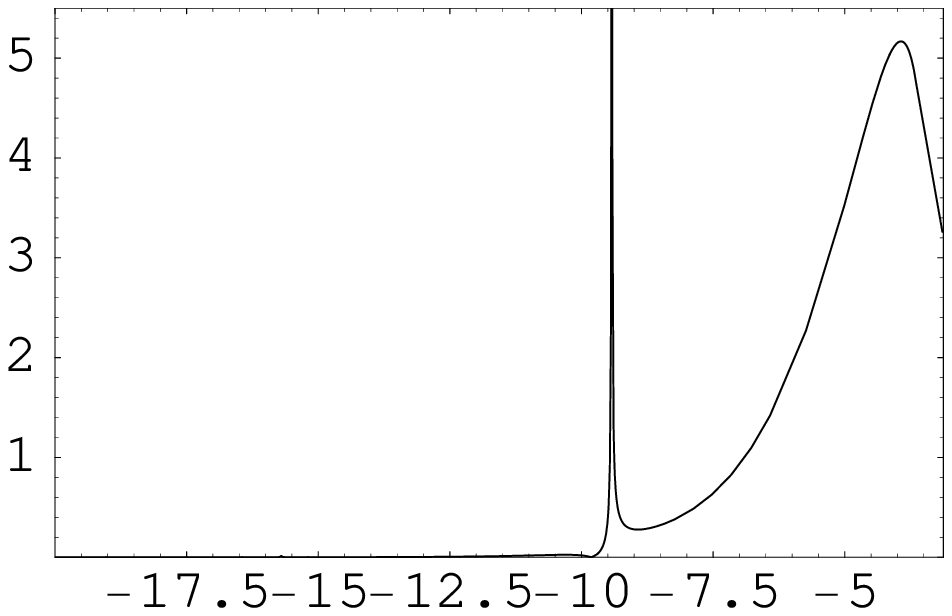}
\includegraphics[width=0.32\textwidth]{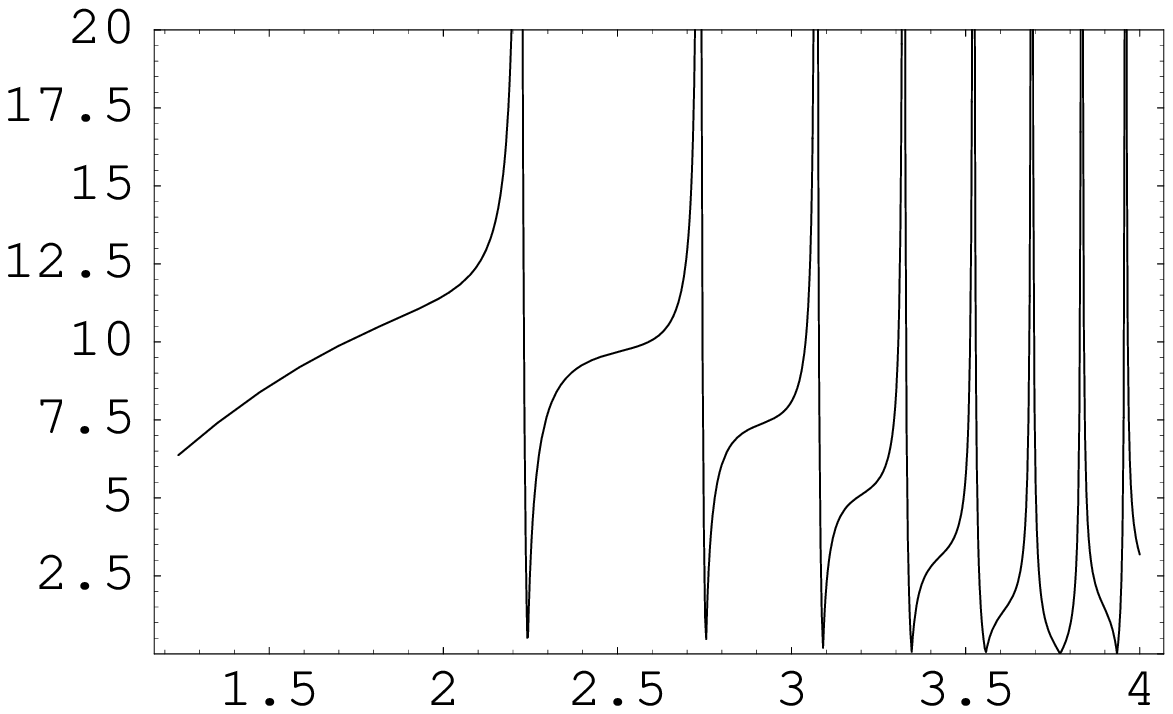}
\\
\hspace{0.5cm}
(a)\hspace{5.5cm}(b)\hspace{5.5cm}(c)
\caption{Upper left graph: The function $\sigma=\sigma(x)$
obtained by integrating Eq.~(\ref{new_EQ_int}) numerically (solid line)
and its approximate expression~(\ref{sig_ap}) with
$\kappa_1=1$ and $\kappa_2=3$  (dashed line).
Upper right graph:
Comparison between the exact solution (solid line)
and the approximate MCE solution~(\ref{chiMCE})
with $\sigma(x)$ evaluated numerically (dashed line) for
$C_+\approx 0.517-0.228\,i$ and $C_-\approx 0.517+0.228\,i$.
Panel~(a) shows the percentage error for the same approximate function
$\sigma$ given in the upper left graph.
Panel~(b) shows the percentage error for the approximate $\chi$ in the upper
right graph for $x<x_1$ and panel~(c) for $x_2<x$.
All plots are for $\tilde\omega=1/2$, $\ell=2$ as in Fig.~\ref{plot1}.}
\label{sigmaOFx}
\end{figure*}
\noindent
The above expression is rather complex.
Hence, we just consider the asymptotic relation between
$\sigma$ and $x$ to the left of the smaller  turning point
(that is, for $x\ll x_1$)
\begin{subequations}
\be
\sigma(x)
\simeq
\frac{\tilde\omega}{\sqrt{D}}\,x
\label{asy_-inf}
\ee
and to the right of the larger one ($x\gg x_2$)
\be
\sqrt{A}\,{\rm e}^{\sigma(x)}-\frac{B}{2\,\sqrt{A}}\,\sigma(x)
\simeq
\tilde\omega\left({\rm e}^{x} +x\right)
\ .
\label{asy_+inf}
\ee
\end{subequations}
\begin{figure*}[t]
\includegraphics[width=0.32\textwidth]{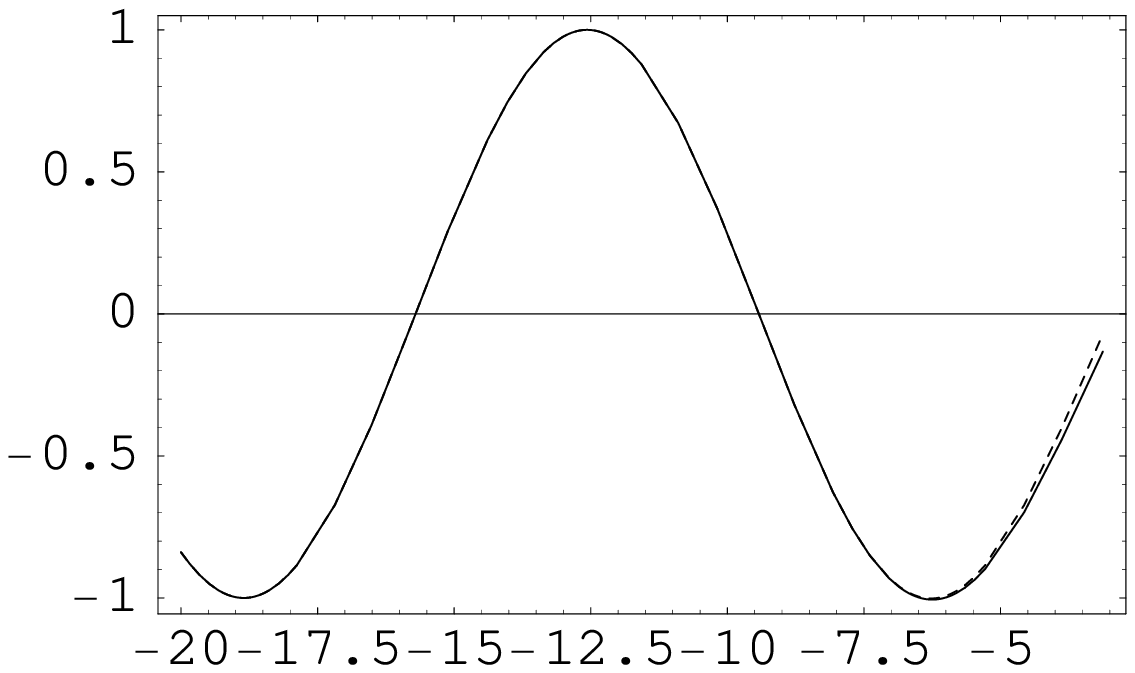}
\includegraphics[width=0.32\textwidth]{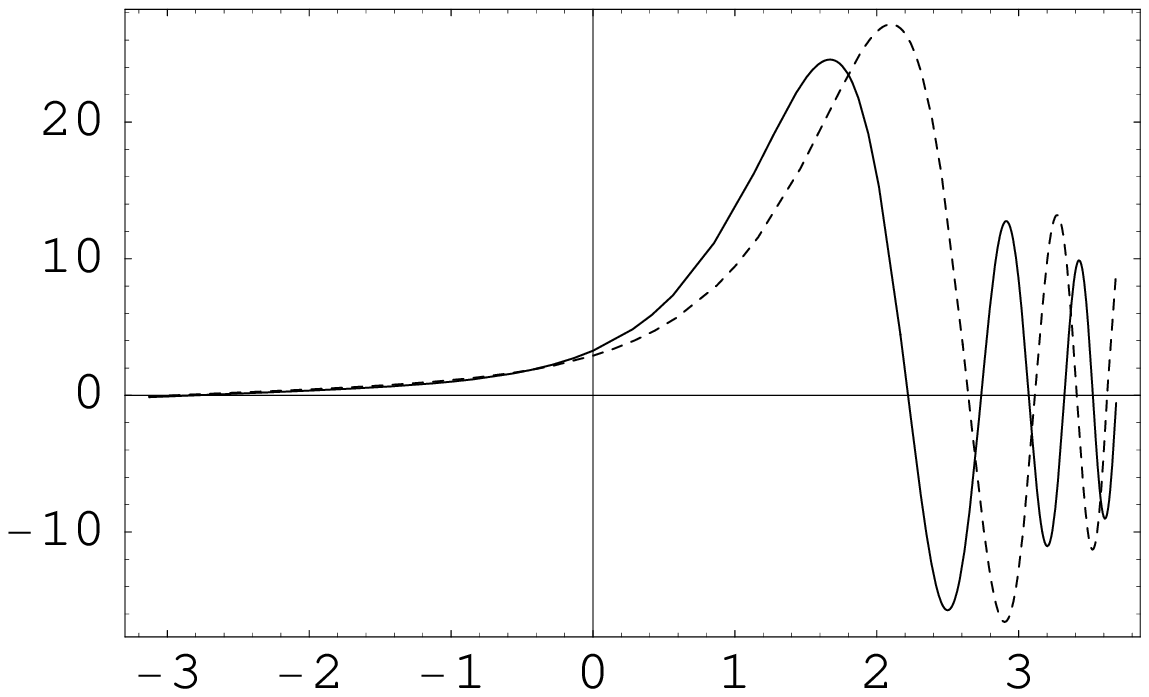}
\includegraphics[width=0.32\textwidth]{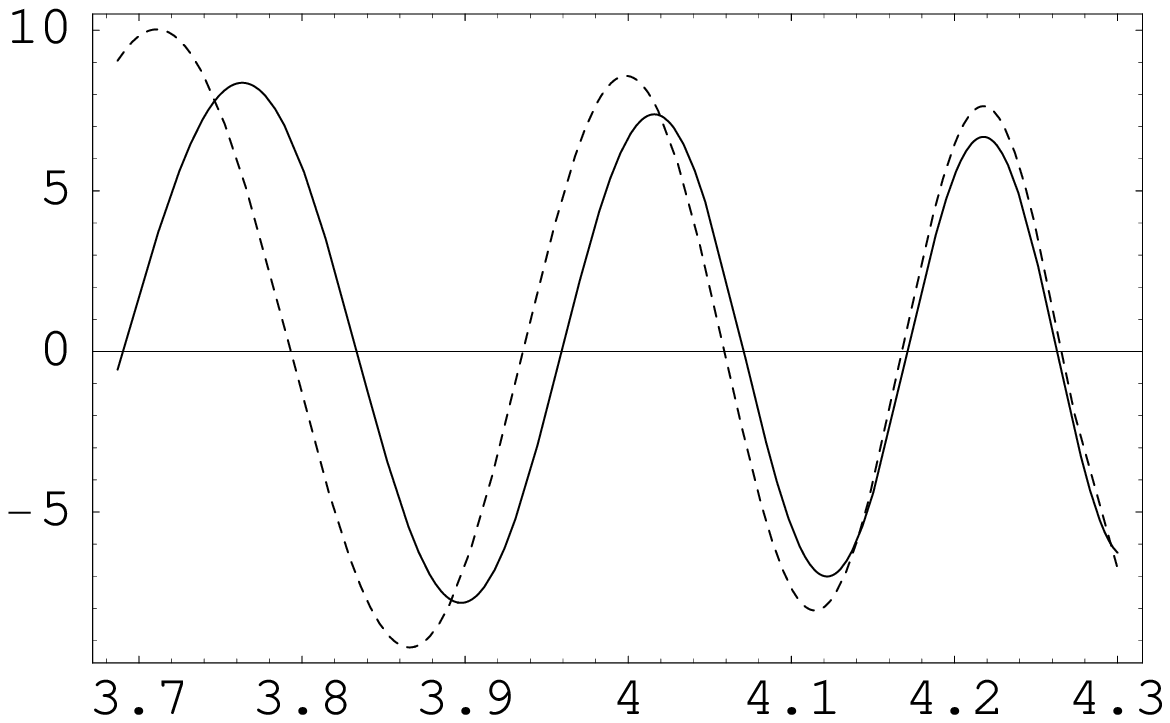}
\\
\hspace{0.5cm}
(a)\hspace{5.5cm}(b)\hspace{5.5cm}(c)
\caption{Comparison between the exact solution (solid line)
and the approximate MCE solution~(\ref{chiMCE})
with $\sigma(x)$ given in Eq.~(\ref{sig_ap}) with
$\kappa_1=1$ and $\kappa_2=3$
(dashed line) for $\tilde\omega=1/2$ and $\ell=2$ (same as in Figs.~\ref{plot1}
and \ref{sigmaOFx}).
Panel (a) shows the result for $x<\kappa_1\,x_1$, panel (b) for
$\kappa_1\,x_1<x<\kappa_2\,x_2$ and panel (c) for $x>\kappa_2\,x_2$.}
\label{final}
\end{figure*}
\par
As an approximate expression for $\sigma(x)$
for $x<\kappa_1\,x_1$ (with $\kappa_1\gtrsim 1$)
we shall then use Eq.~(\ref{asy_-inf}) and the condition that
$\sigma(x_1)=x_1$.
Analogously, for $x>\kappa_2\,x_2$ (with $\kappa_2\gtrsim 1$)
we shall use the approximate solution of Eq.~(\ref{asy_+inf})
to next-to-leading order for large $x$.
Moreover, in the region between $\kappa_1\,x_1$ and $\kappa_2\,x_2$
we shall use an interpolating cubic function.
The reason for introducing two new parameters $\kappa_1$ and
$\kappa_2\gtrsim 1$ is that the asymptotic forms in
Eqs.~(\ref{asy_-inf}) and~(\ref{asy_+inf}) may differ significantly from the correct
$\sigma(x)$ around the turning points $x_1$ and $x_2$ and suitable values of
$\kappa_1$ and $\kappa_2$ usually improve the final result.
To summarize, we have
\be
\sigma\!=\!
\left\{
\begin{array}{ll}
\strut\displaystyle\frac{\tilde\omega}{\sqrt{D}}\,x
+x_1\left(1-\frac{\tilde\omega}{\sqrt{D}}\right)
\ ,
&
x_0<x<\kappa_1 x_1
\\
\\
C_0+C_1\,x+C_2\,x^2+C_3\,x^3
,
&
\kappa_1 x_1<x<\kappa_2 x_2
\\
\\
x+\ln\left(\strut\displaystyle\frac{\tilde\omega}{\sqrt{A}}\right)
\ ,
&
x>\kappa_2 x_2
\ ,
\end{array}
\right.
\label{sig_ap}
\ee
with $\kappa_1$ and $\kappa_2\gtrsim 1$ and
$x_0<\kappa_1\,x_1$ is the value of $x$ at which we wish to
impose the initial conditions for $\chi(x)$ and its derivative~\footnote{Of course,
if we are interested in setting the initial conditions at $x_0\gg x_2$ the expressions
must be adjusted correspondingly.}.
The explicit expressions for the coefficients $C_0$, $C_1$, $C_2$ and $C_3$
are rather cumbersome and will be given in Appendix~\ref{app_cub}.
\par
In the upper left panel of Fig.~\ref{sigmaOFx}, we plot $\sigma(x)$ for the same values of
$\tilde\omega$ and $\ell$ used in Fig.~\ref{plot1}.
The solid line represents the numerical solution of Eq.~(\ref{new_EQ_int})
and the dashed line the approximate expression~(\ref{sig_ap})
with $\kappa_1=1$ and $\kappa_2=3$.
It is clear that these values of $\kappa_1$ and $\kappa_2$ already lead to a very good
approximation for $\sigma$ around the turning points $x_1$ and $x_2$
and that larger values of $\kappa_1$ or $\kappa_2$
would not improve significantly the final result.
In the upper right panel of Fig.~\ref{sigmaOFx}, we compare an MCE approximate
wave-function with the exact solution (solid line) obtained numerically
for the same case.
The MCE solution shown by a dashed line is given by the
expression~(\ref{chiMCE}) with $U(\sigma)$ as in Eq.~(\ref{sol_morse}) and
$\sigma(x)$ determined by solving Eq.~(\ref{new_EQ_int}) numerically.
The coefficients $C_\pm$ are fixed by imposing that the MCE mode
and its derivative equal the corresponding values of the chosen numerical
solution at $x=-20$.
Since the two curves coincide almost everywhere, it is clear that solving
the approximate equation~(\ref{new_EQ_int})
for $\sigma(x)$ is, to all practical extents, equivalent to solving the original
equation~(\ref{exact_EQ}).
\par
In order to further clarify this point, in panel~(a) of Fig.~\ref{sigmaOFx}, we plot the
relative difference (in percent) between the approximate expression~(\ref{sig_ap})
and the exact numerical solution of Eq.~(\ref{new_EQ_int}).
We can see that the error is less than $10\%$ almost everywhere, except
near the zero of $\sigma(x)$.
In the same figure, we also show the percentage error for the approximate
solution shown in the right upper graph of Fig.~\ref{sigmaOFx} for 
$x$ on the left of $x_1$ and on the right of $x_2$.
The latter error remains small, too, except around the zeros of $\chi(x)$
due to the slight phase difference between the two expressions.
\par
Finally, in Fig.~\ref{final}, the dashed lines represent the same MCE solution
with $\sigma(x)$ given by its analytical approximation in Eq.~(\ref{sig_ap})
with $\kappa_1=1$ and $\kappa_2=3$.
This is the fully analytical expression obtained from the application
of the MCE and is still remarkably good, with a more significant error
around the larger turning point ($x_2\approx 1.23$).
From the three graphs in Fig.~\ref{final} and the upper left panel in Fig.~\ref{sigmaOFx},
it clearly appears that such an error develops in the
region between the two turning points, where the discrepancy between
the exact $\sigma(x)$ and the approximate expression~(\ref{sig_ap}) is
indeed larger.
In this respect, let us note that, had we imposed initial conditions
at $x\gg x_2$, the larger error would have occurred around the
smaller turning point ($x_1\approx-3.13$).
Moreover, when we used a linear (or quadratic) interpolating function
for $\kappa_1\,x_1<x<\kappa_2\,x_2$, the accuracy was (expectedly)
worse.
Should one need even better accuracy, a fourth (or higher) order polynomial interpolation
between $\kappa_1\,x_1$ and $\kappa_2\,x_2$ could be used instead. 
\section{Conclusions}
\label{sC}
We have applied the method of comparison equations to a specific
problem of scalar field dynamics on the Schwarzschild background.
We have considered modes with energy lower than the peak of the
effective potential in the corresponding wave equation, so that there
are two zeros (turning points).
The comparison function has then been chosen of the Morse form
with an argument which we have estimated analytically as well.
This procedure was shown to be able to produced a fully analytical
approximation of the wave-functions with good accuracy over the
whole domain outside the Schwarzschild horizon.
\par
Let us now make a few remarks.
Firstly, the application of the MCE is not as straightforward as
the standard WKB approximation.
In fact, the MCE usually requires solving some technical tasks specific to
the problem at hand in order to obtain the argument $\sigma=\sigma(x)$
of the known function $U=U(\sigma)$.
This aspect has already been analyzed in the simpler situation of cosmological
perturbations for which one needs consider just one
turning point~\cite{MCEcosmo}.
In the present case, with two turning points, all expressions become
more involved.
For instance, the best approximation for $\sigma(x)$ we found convenient to display
in Eq.~(\ref{sig_ap}) makes use of a cubic interpolation over
the region containing the two turning points and is given in terms of
two arbitrary parameters $\kappa_1$ and $\kappa_2$ which
can be fixed in order to minimize the error (and explicit expressions for the
coefficients $C_0,\ldots,C_3$ are given in Appendix~\ref{app_cub}).
Of course, one could employ higher order interpolating functions
and further improve the result at the price of complexity if better
accuracy is needed. 
\par
Finally, our aim for the present work was mainly to test the effectiveness
of the MCE to produce approximate wave modes on black hole backgrounds.
For this reason, no discussion of gauge fields, gravitational waves or spinors
has been explicitly included.
We in fact plan to apply the MCE also to such higher spin fields and further extend
the method to study related problems,
such as the quasi-normal modes~\cite{nollert} and grey-body factors~\cite{hawking}
of radiating black holes (also in the brane-world~\cite{bw}, given the relevance
of such objects in future collider experiments~\cite{lhc}).
%
%
%
%
%
%
\appendix
\section{Turning Points}
\label{app_TPs}
If we define $w={\rm e}^{x}$, we see that $\omega^2(x)=0$
for the exact frequency in Eq.~(\ref{freq_ex}) becomes
the quartic polynomial equation 
\begin{subequations}
\be
w^4 + 4\,w^3 + b\,w^2 + c\,w + 1 = 0
\ ,
\ee
with
\be
b
&\!=\!&
6 - \frac{1}{4\,{\tilde\omega }^2} - \frac{\ell\left(\ell+1\right)}{{\tilde\omega }^2}
\\
c
&\!=\!&
4 - \frac{1}{2\,{\tilde\omega }^2} - \frac{\ell\left(\ell+1\right)}{{\tilde\omega }^2}
\ .
\ee
\end{subequations}
Its four roots  can be written as
\begin{subequations}
\be
w_i
&\!=\!&
-1
+\frac{\epsilon_i}{2}
\sqrt{
4-\frac{2}{3}\,b
+\frac{\gamma}{3}
}
\nonumber
\\
&&
-\frac{\theta_i}{2}
\sqrt{
8 - \frac{4}{3}\,b
+\epsilon_i\frac{4\,b -16 - 2\,c}{{\sqrt{4-\frac{2}{3}\,b
+\frac{\gamma}{3}
}}}
-\frac{\gamma}{3}
}
\ ,
\ee
where $i=1,\ldots,4$ and
\be
\!\!\!\!\!\!\!\!\!\!\!\!\!\!\!\!
\alpha
&\!=\!&
12 + b^2 - 12\,c
\\
\!\!\!\!\!\!\!\!\!\!\!\!\!\!\!\!
\beta
&\!=\!&
2\,b^3 - 36\,b\,c + 27\,c^2 + 432 - 72\,b
\\
\!\!\!\!\!\!\!\!\!\!\!\!\!\!\!\!
\gamma
&\!=\!&
\frac{2^{\frac{1}{3}}\,\alpha}{\left(\beta+\sqrt{\beta^2-4\,\alpha^3}\right)^{\frac{1}{3}}}
+\frac{\left(\beta+\sqrt{\beta^2-4\,\alpha^3}\right)^{\frac{1}{3}}}{2^{\frac{1}{3}}}
\ .
\ee
\end{subequations}
The coefficients $\epsilon_1=\epsilon_2=-\epsilon_3=-\epsilon_4=
\theta_1=-\theta_2=\theta_3=-\theta_4=1$, and the values of $\tilde\omega$ and $\ell$
determine the signs of the roots.
There can be at most two positive roots, in which case, say,
$w_3\le w_4\le 0\le w_1={\rm e}^{x_1}\le w_2={\rm e}^{x_2}$, and only
$x_1$ and $x_2$ can be valid turning points.
The critical value of $\tilde\omega$ in Eq.~(\ref{om_c}) is then determined by
the condition $0\le w_1=w_2$.
\section{Evaluation of $\xi$}
\label{app_xi}
In order to evaluate the integral in Eq.~(\ref{third_cond}), we start from the general
relation (again with $w={\rm e}^x$)
\begin{widetext}
\be
\frac{1}{\tilde\omega}
\int_{x_1}^{\ln(w)}\sqrt{-\,\omega^2(y)}\,{\d}\,y
&\!=\!&
\sqrt{\frac{(w-w_3) (w-w_1) (w-w_2)}{w-w_4}}
+\frac{1}{\sqrt{(w_1-w_3)(w_4-w_2)}}
\nonumber
\\
&&
\times
\left\{
(w_3-w_1)(w_4-w_2)\,{\rm E}(z;W)
-(w_4-w_1)(w_4-w_2)\,{\rm F}(z;W)
\phantom{\frac{A}{B}}
\right.
\nonumber
\\
&&
\phantom{\times\ \ }
\left.
+(w_4-w_1)(w_3+w_4+w_1+w_2+2)\,
{\rm \Pi}\!\left[\frac{w_1-w_2}{w_4-w_2};z;W\right]
\right.
\nonumber
\\
&&
\phantom{\times\ \ }
\left.
-2\,(w_4-w_1)(w_3+1)(w_2+1)\,{\rm \Pi}\!
\left[\frac{(w_4+1)(w_1-w_2)}{(w_1+1)(w_4-w_2)};z;W\right]
\right.
\nonumber
\\
&&
\phantom{\times\ \ }
\left.
+2\,(w_4-w_1)w_3\,w_2\,{\rm \Pi}\!
\left[\frac{w_4\,(w_1-w_2)}{w_1\,(w_4-w_2)};z;W\right]
\right\}
\ ,
\label{xi_gen}
\ee
\end{widetext}
where E, F and ${\rm \Pi}$ are elliptic functions~\cite{abramowitz},
\be
z={\rm ArcSin}\!\left[\sqrt{\frac{(w-w_1)(w_4-w_2)}{(w-w_4)(w_1-w_2)}}\right]
\ ,
\ee
\be
W=\frac{(w_3-w_4)(w_1-w_2)}{(w_3-w_1)(w_4-w_2)}
\ ,
\ee
and $w_i={\rm e}^{x_i}$ are the roots of $\omega^2(x)$
as given in the previous Appendix.
The above expression evaluated at $w=w_2$ yields
\be
\!\!\!\!\!\!\!\!
&&
\xi=
\frac{\tilde\omega}{\sqrt{(w_1-w_3)(w_4-w_2)}}
\nonumber
\\
\!\!\!\!\!\!\!\!
&&
\times
\left\{(w_3-w_1)(w_4-w_3)\,{\rm E}(W)
\phantom{\frac{A}{B}}
\nonumber
\right.
\\
\!\!\!\!\!\!\!\!
&&
\phantom{\times\ \ }
\left.
+(w_4-w_1)\,(w_2-w_4)\,{\rm K}(W)
\right.
\nonumber
\\
\!\!\!\!\!\!\!\!
&&
\left.
+(w_4-w_1)\,(w_3+w_4+w_1+w_2+2)\,{\rm \Pi}\!
\left[\frac{w_1-w_2}{w_4-w_2};W\right]
\right.
\nonumber
\\
\!\!\!\!\!\!\!\!
&&
\left.
-2 (w_3+1)(w_2+1)(w_4-w_1)\,{\rm \Pi}\!
\left[\frac{(w_4+1)(w_1-w_2)}{(w_1+1)(w_4-w_2)};\!W\!\right]
\right.
\nonumber
\\
\!\!\!\!\!\!\!\!
&&
\phantom{\times\ \ }
\left.
+2\,w_3\, w_2\,(w_4-w_1)\,{\rm \Pi}\!
\left[\frac{w_4\,(w_1-w_2)}{w_1\,(w_4-w_2)};W\right]
\right\}
\ ,
\label{xi}
\ee
in which we used E$(\pi/2;q)=$E$(q)$,
F$(\pi/2;q)=$K$(q)$ and
$\Pi(p;\pi/2;q)=\Pi(p,q)$~\cite{abramowitz}.
\section{Cubic interpolation}
\label{app_cub}
In Section~\ref{sMCE_BH}, we use the analytic approximation for $\sigma(x)$
in Eq.~(\ref{sig_ap}) which involves the cubic
interpolation
\be
\sigma=C_0+C_1\,x+C_2\,x^2+C_3\,x^3
\ ,
\ee
for $\kappa_1\,x_1<x<\kappa_2\,x_2$.
The above coefficients $C_i$, with $i=0,\ldots,3$, can be easily expressed
in terms of the parameters
$A$, $B$ and $D$
for the Morse potential~(\ref{comp_freq}) as
\begin{subequations}
\begin{widetext}
\be
\!\!\!\!\!\!\!\!\!\!\!\!\!\!\!\!\!\!\!\!\!\!\!\!
&&
C_0=
{\kappa_1}\,{x_1}\,
\frac{\left(2\,{\sqrt{D}-1} \right) \,{{\kappa_2}}^2\,{{x_2}}^2\,
\left( {\kappa_2}\,{x_2} - {\kappa_1}\,{x_1} \right)
-2\,{\sqrt{D}}\,\ln \left(2\,{\sqrt{A}}\right)
{\kappa_1}\,{x_1}
\left(3\,{\kappa_2}\,{x_2}- {\kappa_1}\,{x_1} \right)}
{2\,{\sqrt{D}}\,{\left( {\kappa_2}\,{x_2} - {\kappa_1}\,{x_1} \right) }^3}
\\
\!\!\!\!\!\!\!\!\!\!\!\!\!\!\!\!\!\!\!\!\!\!\!\!
&&
C_1=
\frac{\left( {\kappa_2}\,{x_2} - {\kappa_1}\,{x_1} \right)
\left[
{\kappa_2}\,{x_2}\,\left( 2\,{\kappa_1}\,{x_1} + {\kappa_2}\,{x_2} \right)
-2\,{\sqrt{D}}\,{\kappa_1}\,{x_1}\,
\left( 4\,{\kappa_2}\,{x_2} -{\kappa_1}\,{x_1} \right)
\right] 
+12\,{\sqrt{D}}\,\ln \left(2\,{\sqrt{A}}\right)\,{\kappa_1}\,{\kappa_2}\,{x_1}\,{x_2} 
}
{2\,{\sqrt{D}}\,{\left( {\kappa_2}\,{x_2} - {\kappa_1}\,{x_1} \right) }^3}
\\
\!\!\!\!\!\!\!\!\!\!\!\!\!\!\!\!\!\!\!\!\!\!\!\!
&&
C_2=
\frac{6\,{\sqrt{D}}\,\ln \left(2\,{\sqrt{A}}\right)
\left( {\kappa_1}\,{x_1} + {\kappa_2}\,{x_2} \right)  +
\left(2\,{\sqrt{D}}-1 \right)
\left({{\kappa_1}}^2\,{{x_1}}^2 + {\kappa_1}\,{\kappa_2}\,{x_1}\,{x_2}
-2\,{{\kappa_2}}^2\,{{x_2}}^2 \right) }{2\,{\sqrt{D}}
{\left( {\kappa_1}\,{x_1} - {\kappa_2}\,{x_2} \right) }^3}
\\
\!\!\!\!\!\!\!\!\!\!\!\!\!\!\!\!\!\!\!\!\!\!\!\!
&&
C_3=
\frac{4\,{\sqrt{D}}\,\ln \left(2\,{\sqrt{A}}\right)
-\left(2\,{\sqrt{D}}-1\right) \,\left( {\kappa_2}\,{x_2} - {\kappa_1}\,{x_1} \right) }
{2\,{\sqrt{D}}\,{\left( {\kappa_2}\,{x_2} - {\kappa_1}\,{x_1} \right) }^3}
\ .
\ee
\end{widetext}
\end{subequations}
Finally, one can express everything in terms of the original parameters
$\tilde\omega$ and $\ell$ by making use of Eqs.~(\ref{coeff_A})-(\ref{coeff_D}).

\end{document}